\documentclass[twocolumn,showpacs,superscriptaddress,groupedaddress,nobibnotes,prb]{revtex4-1}
\usepackage[]{graphicx}
\usepackage{setspace}
\usepackage{tocloft}
\usepackage{amsfonts}
\usepackage{amsmath}
\usepackage{amsbsy}
\usepackage{amstext}
\usepackage{amssymb}
\usepackage{fixmath}

\cftpagenumbersoff{figure}
\cftpagenumbersoff{table} 

\newcommand{\eps}{\varepsilon}
\newcommand{\EQE}[1]{\eta_{\text{EQE}#1}}
\newcommand{\out}[1]{\eta_{\text{out}#1}}
\newcommand{\mate}{{\text{e}}}

\begin{document} 
\title{Enhanced light emission from top-emitting organic light-emitting diodes by optimizing surface plasmon polariton losses} 

\author{Cornelius Fuchs}
\email{cornelius.fuchs@iapp.de}
\affiliation{Technische Universit\"at Dresden, Institut f\"ur Angewandte Photophysik, George-B\"ahr-Stra{\ss}e 1, Dresden, Germany, 01069}

\author{Paul-Anton Will}
\affiliation{Technische Universit\"at Dresden, Institut f\"ur Angewandte Photophysik, George-B\"ahr-Stra{\ss}e 1, Dresden, Germany, 01069}

\author{Martin Wieczorek}
\affiliation{Technische Universit\"at Dresden, Institut f\"ur Angewandte Photophysik, George-B\"ahr-Stra{\ss}e 1, Dresden, Germany, 01069}

\author{Malte C. Gather}
\altaffiliation[Now at: ]{SUPA, School of Physics and Astronomy, University of St Andrews, North Haugh, St Andrews, Scotland, UK, KY16 9SS}
\affiliation{Technische Universit\"at Dresden, Institut f\"ur Angewandte Photophysik, George-B\"ahr-Stra{\ss}e 1, Dresden, Germany, 01069}

\author{Simone Hofmann}
\affiliation{Technische Universit\"at Dresden, Institut f\"ur Angewandte Photophysik, George-B\"ahr-Stra{\ss}e 1, Dresden, Germany, 01069}

\author{Sebastian Reineke}
\affiliation{Technische Universit\"at Dresden, Institut f\"ur Angewandte Photophysik, George-B\"ahr-Stra{\ss}e 1, Dresden, Germany, 01069}

\author{Karl Leo}
\homepage{http://www.iapp.de}
\affiliation{Technische Universit\"at Dresden, Institut f\"ur Angewandte Photophysik, George-B\"ahr-Stra{\ss}e 1, Dresden, Germany, 01069}

\author{Reinhard Scholz}
\affiliation{Technische Universit\"at Dresden, Institut f\"ur Angewandte Photophysik, George-B\"ahr-Stra{\ss}e 1, Dresden, Germany, 01069}


\begin{abstract}
We demonstrate enhanced light extraction for monochrome top-emitting organic light-emitting diodes (OLEDs).
The enhancement by a factor of 1.2 compared to a reference sample is caused by the use of a hole transport layer (HTL) material possessing a low refractive index ($\sim 1.52$).
The low refractive index reduces the in-plane wave vector of the surface plasmon polariton (SPP) excited at the interface between the bottom opaque metallic electrode (anode) and the HTL.
The shift of the SPP dispersion relation decreases the power dissipated into lost evanescent excitations and thus increases the outcoupling efficiency, although the SPP remains constant in intensity.
The proposed method is suitable for emitter materials owning isotropic orientation of the transition dipole moments as well as anisotropic, preferentially horizontal orientation, resulting in comparable enhancement factors.
Furthermore, for sufficiently low refractive indices of the HTL material, the SPP can be modeled as a propagating plane wave within other organic materials in the optical microcavity.
Thus, by applying further extraction methods, such as micro lenses or Bragg gratings, it would become feasible to obtain even higher enhancements of the light extraction.
\end{abstract}

\pacs{78.60.Fi, 78.66.Qn, 85.60.Bt, 85.60.Jb, 73.20.Mf}



\maketitle 

\section{Introduction}
Organic light-emitting diodes (OLEDs) are already applied as lighting or display technology today.\cite{Reineke,2011_Mladenovski_Integrated_optical_model_for_OLED}
However, the research for more efficient devices is widespread.\cite{Reineke,Thomschke2012,2013_Xiang_study_enhancements_phos_microcav_OLEDs,2013_Reineke_white_Review}
Especially top-emitting OLEDs are of great interest, as these do not radiate through a transparent substrate and therefore allow for a more flexible choice of substrate materials.\cite{1996_Bulovic_Transparent_top_emitting_OLEDs_Nature,Hofmann:11,Thomschke2012}
The radiometric efficiency of an OLED is given as the external quantum efficiency (EQE) $\EQE{}$, relating the number of extracted photons to the number of injected charges.\cite{2006_BOOK_Dakin_Handbook_of_OptElec}
Increasing the EQE remains an important goal in OLED related research.
The EQE can be divided into a product of three efficiencies $\EQE{} = \gamma \, \eta_\text{IQE} \, \out{}$.\cite{Patel}
In this relation, $\gamma$ represents the electrical efficiency, accounts for electrical losses and charge carrier balance.
The internal quantum efficiency denoted by $\eta_\text{IQE}$ characterizes the efficiency of charge to photon conversion.
Both quantities are close to unity for common \textit{pin}-OLEDs\cite{Pfeiffer2003,Walzer2007} incorporating phosphorescent emitters.\cite{1998_Baldo_Eff_phosph_OLEDs,Adachi,2010_Kwon_Irppy3_and_deriv_PL-yield,2014_Kim_Horizont_phosph_emitter_OLEDs}
However, due to the high refractive indices of the organic semiconductor materials causing total internal reflection (TIR),\cite{Barnes,Patel} the outcoupling efficiency $\out{}$ is typically limited to values below 30\,\%.
Therefore, the focus in OLED related research is increasingly on $\out{}$.
The 30\,\% limit takes into account the enhanced power dissipation into resonances of the thin optical microcavity.\cite{Barnes,FurnoII}
For top-emitting OLEDs, the optical losses can be divided into contributions caused by the finite extent of the optical microcavity, i.e. optical wave guide (WG) modes, losses from absorption, and excitations of surface plasmon polariton (SPP) modes at metal interfaces.\cite{Meerheim}
Either the OLED suffers from large losses due to the SPP mode, which is excited at the metallic opaque back reflector, or, if the distance between organic emitter and the metal is increased, another WG mode as source of loss is introduced, while the SPP excitation is strongly reduced.\cite{Meerheim}
Typically, even though the experimentalist is free to ponder between different types of loss channels, known OLED layouts cannot exceed the limit mentioned above.
For top-emitting OLEDs without any further outcoupling techniques, the first order optical microcavity incorporating a large SPP contribution was found to be more efficient than optical microcavities exhibiting higher order WG modes.\cite{2010_Hofmann_APL_top_1st_to_3rd_order_and_metal_variation,Meerheim}
Lately, anisotropic emitter materials showing a preferred spatial orientation of their transition dipole moment received great attention.\cite{2012_Bruetting_PSS_Device_efficiency_of_OLEDs_Progress_by_improved_light_outcoupling,2014_Kim_Horizont_phosph_emitter_OLEDs,Wasey}
Using such materials, the efficiency limit can be increased compared to isotropic emitters.\cite{2014_Kim_Horizont_phosph_emitter_OLEDs,Wasey}
For spatial orientations of common phosphorescent emitter materials the enhancement was found to be about 15\,\% for optimized devices, as the contribution of plasmonic losses for the first order device are reduced.\cite{2014_Kim_Horizont_phosph_emitter_OLEDs}
Recently, for nearly completely horizontal alignment, the enhancement was found to be up to 50\,\%\cite{2014_Mayr_Emitter_0.08_orientation_TADF}, in accordance with simulations\cite{Schmidt}.
To extract the remaining power lost into SPP excitations, scattering approaches have been proposed\cite{Bi,2012_Bruetting_PSS_Device_efficiency_of_OLEDs_Progress_by_improved_light_outcoupling,2011_Frischeisen_many_tech_SPP_out_OLED}, or methods involving complex microlens structures of high refractive index.\cite{2012_Bruetting_PSS_Device_efficiency_of_OLEDs_Progress_by_improved_light_outcoupling,2011_Frischeisen_many_tech_SPP_out_OLED,Scholz}

Here, we propose a method to increase the outcoupling efficiency $\out{}$ of OLEDs by modifying the properties of the SPP excitations of the simple planar optical cavity.
We avoid power dissipation into evanescent modes by using Poly(3,4-ethylenedioxythiophene)-poly(styrenesulfonate) (PEDOT:PSS) as doped organic charge transport materials with a low refractive index.
The reduction of the average refractive index by about 14\,\% results in an average increase of the outcoupling efficiency of about 20\,\% regardless of whether the emitter exhibits anisotropic or isotropic orientation of the transition dipole moments, as the SPP excitation itself is retained.
These results, initially obtained from optical simulations, are experimentally verified for optimized first-order green phosphorescent top-emitting OLEDs with emitters having isotropic or preferential horizontal orientation, respectively.

Several studies have successfully implemented low refractive index materials such as PEDOT:PSS into OLEDs.
These works focused on the use of low refractive index materials or PEDOT:PSS as electrode\cite{2011_Cai_PEDOT_in_OLEDs_but_as_anode_and_bottom,Yoo,2014_enhanced_light_outcoupling_from_PEDOT_elec_comp_ITO}, hole transport layers\cite{Koehnen} or low refractive index spacers outside the electrical active microcavity\cite{2001_Tsutsui_AdvMater_low_index_substrate} in bottom-emitting OLEDs, or low refractive index photonic crystal layers.\cite{Sun}
For these stratified planar bottom-emitting OLEDs, the enhancements of the EQE compared to common materials (i.e. indium tin oxide electrodes) resulted from the drastic reduction of the microcavity size or mediation of the properties of the wave guide.
So in these works OLEDs of different optical orders were compared, instead of OLEDs with similar thickness of the micro cavity.
Furthermore, in a purely theoretical study, Smith and Barnes\cite{Smith} investigated the effect of reducing the refractive index of the entire organic layers within the optical microcavity for bottom and top-emitting geometries.
They found great potential for improvement, because with decreasing refractive index the resonant enhancement of non-radiative modes is reduced.

In contrast, the focus of this work is the enhanced light harvesting of the already optimized first order top-emitting OLEDs and experimental verification of the theoretical predictions.
For these devices, the optical thickness of the microcavity layers can not be reduced to obtain higher optical outcoupling efficiency\cite{Meerheim}, neither by reducing the layer thickness, nor by reducing the refractive index.
Instead, we mitigate the losses from surface plasmon contributions, increasing the boundary of the theoretical limit of outcoupling efficiency for planar OLEDs.
In addition, a detailed discussion of the optical effects is given.
The theoretical discussion indicates that in the best case the modified SPP can be modeled by a non-evanescent plane wave within the organic layers.
Thus, it becomes an accessible target for further macroscopic\cite{2007_Yang_top_OLEDs_w_structure_4_outcoupling,Thomschke2012} or microscopic\cite{Bi,2014_Schwab_Fuchs_Opt_Exp} outcoupling techniques.

\section{Devices and materials}\label{sec_materials}
\begin{figure*}
\onecolumngrid
\centering
\begin{minipage}{0.9\textwidth}
\includegraphics[width=\textwidth]{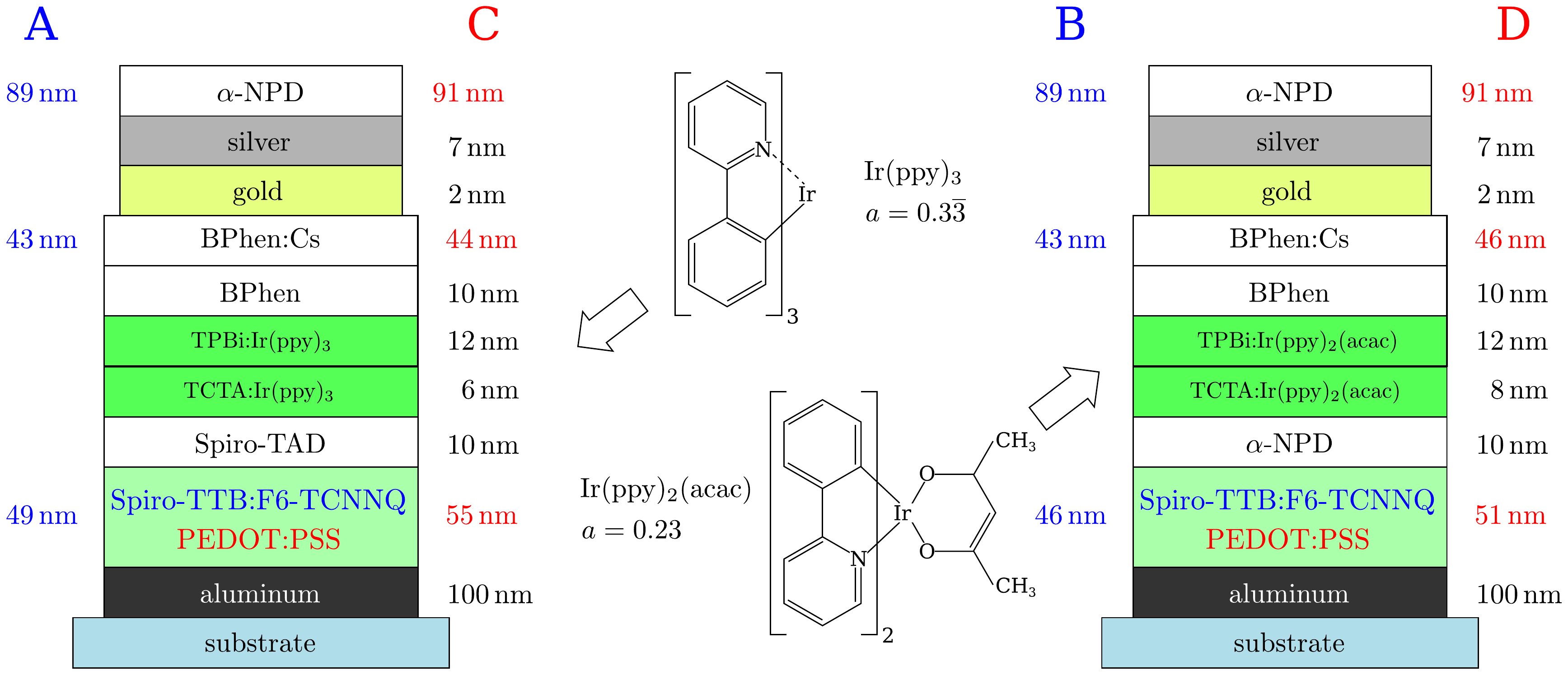}
\end{minipage}
\caption{Scheme of the fabricated devices A to D. The devices are distinguished by their phosphorescent emitter molecules, which show either isotropic ($a=0.3\overline{3}$, Ir(ppy)$_3$, devices A and B) orientation of their transition dipole moment, or anisotropic orientation (i.e. $a=0.23$, Ir(ppy)$_2$(acac), devices C and D). Devices A and C are reference devices with high refractive index HTL, i.e. Spiro-TTB:F6-TCNNQ, devices B and D have a low refractive index PEDOT:PSS HTL.}\label{fig_devices}
\end{figure*}

This report investigates plain colored top-emitting OLEDs.
All these devices follow the \textit{pin}-concept using doped charge transport layers enclosing intrinsic charge blocking and emission layers.\cite{Pfeiffer2003,Walzer2007}
The thicknesses of the transport and organic capping layer were obtained by optimization of the optical outcoupling efficiency from optical simulations.
Hereby the fixed dimensions of the intrinsic layers of the OLED were taken into account.
The thicknesses of the intrinsic layers were obtained in order to achieve the highest internal quantum efficiencies.\cite{2014_Murawski_Cavity_Design_Roll_off,FurnoII}
To guide the reader, the layer sequences of devices A to D and their distinctive features are illustrated in Fig.~\ref{fig_devices}.
The OLEDs were manufactured onto glass substrates, upon which the opaque bottom aluminum electrode (anode, 100\,nm) was deposited.
Next, the hole transport layer (HTL) was fabricated.
For devices A and C, both serving as reference devices, the HTL was evaporated using 2,2',7,7'-tetra(N,N-ditolyl)amino-9,9-spiro-bifluorene (Spiro-TTB) doped with 4\,wt.\% of 2,2 -(perfluoro-naphthalene-2,6-diylidene)dimalononitrile (F6-TCNNQ), where the thickness was 49\,nm for device A and 46\,nm for device C.
The HTLs of devices B and D were prepared from PEDOT:PSS (Clevios P AI4083) with a thickness of 55\,nm for Device B and 51\,nm for device D.
Among the two different HTL materials used in this study, Spiro-TTB:F6-TCNNQ has a high refractive index (average value of $n_\text{high} = 1.77$ over the emission spectrum of the emitter molecules), whereas PEDOT:PSS possesses a low average refractive index $n_\text{low}=1.52$.
The intrinsic emission units were deposited onto the HTLs for each device.
They consisted of an electron blocking layer (EBL), a double emission layer (EML), and a hole blocking layer (HBL).
The EBL was fabricated from 2,2',7,7'-Tetrakis-(N,N-diphenylamino)-9,9'-spirobifluorene (Spiro-TAD, 10\,nm), for devices A and B.
For devices C and D the EBL was prepared from N,N'-Di(naphthalen-1-yl)-N,N'-diphenyl-benzidine ($\alpha$-NPD, 10\,nm).
The EML of device A and C were produced from 4,4',4''-tris(carbazol-9-yl)-triphenylamine (TCTA, 6\,nm) as one host material, and 2,2',2''-(1,3,5-Phenylen)tris(1-phenyl-1H-benzimidazole) (TPBi, 12\,nm) as a second EML.
As emission material for devices A and C the phosphorescent green emitter Tris(2-phenylpyridine) iridium(III) (Ir(ppy)$_3$) was doped with a concentration of 8\,wt\% into these hosts.
The double EML host materials for devices C and D are the same as for devices A and B, but the optimized thicknesses were adjusted to 8\,nm of TCTA and 12\,nm of TPBi.
In the devices C and D the phosphorescent green emitter dopant bis(2-phenylpyridine)(acetylacetonate) iridium(III) (Ir(ppy)$_2$(acac)) doped with 8\,wt\% into the matrix was used.
In all devices the HBL was produced from 4,7-diphenyl-1,10-phenanthroline (BPhen, 10\,nm).

The choice of the different green emitting molecules was motivated by their different alignment of their transition dipole moments.
Ir(ppy)$_3$ is known to have an isotropic dipole distribution\cite{2012_Liehm_orientation_APL_Ir3_Ir2_green}, where in contrast Ir(ppy)$_2$(acac) shows preferential horizontal (preferential parallel to the $x$-, resp. $y$-axis) alignment.\cite{2012_Liehm_orientation_APL_Ir3_Ir2_green}
To complete the OLED stack, an electron transport layer (ETL) of BPhen doped with cesium, a transparent metal electrode (cathode) consisting of 2\,nm gold and 7\,nm silver, and an organic capping layer (CL) of $\alpha$-NPD were deposited.
For device A and B the thicknesses of the ETL and CL were 43\,nm and 89\,nm, and for device C and D the ETL thickness was 44\,nm and 46\,nm respectively, and a CL had a thickness of 91\,nm.

\section{Theoretical concept}
\subsection{Shift of the SPP excited at the dielectric/opaque metal interface}
Every metal/dielectric interface supports SPPs, arising from the coupling between surface charge oscillations and the electromagnetic field.
In OLEDs, SPPs can be excited due to the close proximity between the metal/dielectric interface and the radiation source.
Commonly at least the opaque bottom electrode in top-emitting OLEDs is produced from metal, because of the convenient large broadband reflectivity and simplicity in fabrication\cite{2010_Hofmann_APL_top_1st_to_3rd_order_and_metal_variation,Hofmann:11,2014_Cao_transparent_electrode_review_OLED} although there have been attempts to fabricate semi-transparent top contacts from sputtered transparent conductive oxides (TCOs) or polymers.\cite{2014_Cao_transparent_electrode_review_OLED,2010_Chen_Recent_Dev_TOLEDs}

\begin{figure}
\centering
\begin{minipage}{0.5\textwidth}
\includegraphics[width=\textwidth]{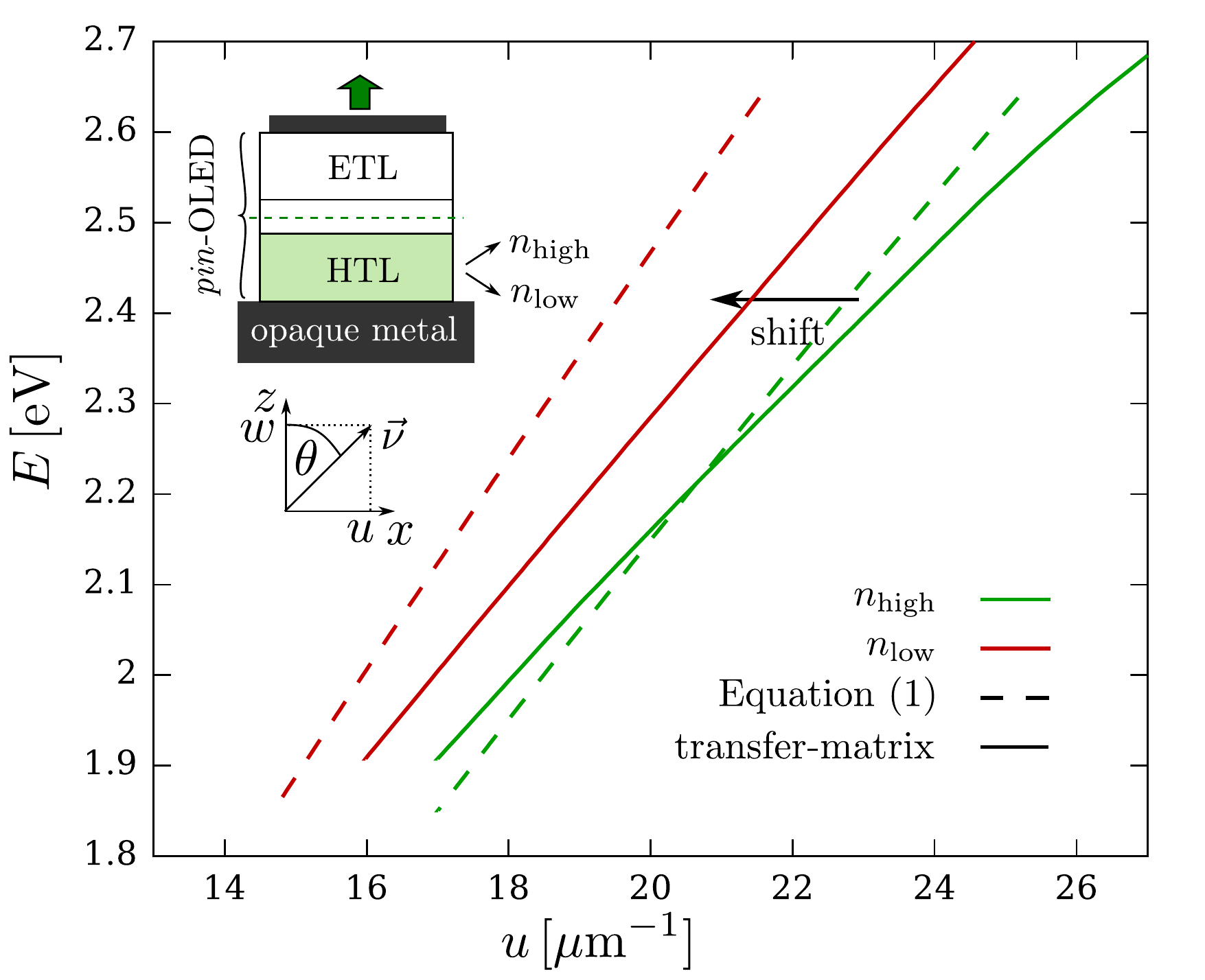}
\end{minipage}
\caption{Calculated dispersion relations of oSPP modes. The calculations are carried out either in the approximation of Eq.~(\ref{eq_disp_oSPP}), or by taking into account the complete OLED optical microcavity by using the transfer-matrix approach. For both methods a shift of the oSPP mode towards smaller in-plane wavenumbers $u$ is observed for a decreased refractive index of the dielectric medium adjacent to the opaque metal.}\label{fig_oSPP_disp_shift}
\end{figure}

For the theoretical discussion, the definition of the coordinate system in reference to the OLED is shown in the inset of Fig.~\ref{fig_oSPP_disp_shift}.
To outline the properties of the SPP within OLED microcavities at first a brief discussion for a simplified interface of two adjacent media is given.
The electromagnetic field of an SPP mode, modeled by a p-polarized time-harmonic electromagnetic plane wave with its magnetic field $H_\text{y} = H_{0,\,\text{y}} \, \text{e}^{\imath \left( u \, x + w \, z \right)}$, is localized by an exponentially decaying envelope around the metal/dielectric interface.
The coordinate system is laid out such that the plane of incidence, defined by the wave vector $\vec{\nu}$ of the electromagnetic plane wave and the $z$-direction is given by $y=0$.
Thus, the y-component of the wave vector is zero.
Furthermore, the interfaces between the stratified media are located at planes denoted by $z=\text{const.}$\,.
The localization of the electromagnetic field of the simple SPP can be seen from the dispersion relation excited at an interface at $z=0$ between a metal and a dielectric medium with dielectric functions $\eps_\text{m}$ and $\eps_\text{d}$ \cite{Maier}
\begin{equation}\label{eq_disp_oSPP}
E = \hbar \, c \, u \sqrt{\frac{\eps_\text{m} + \eps_\text{d}}{\eps_\text{m} \, \eps_\text{d}}}\, .
\end{equation}
Here, $E$ denotes the energy, $u$ the in-plane component of the wave vector, $c$ identifies the speed of light and $\hbar$ the Planck constant.
The localization of the SPP is understood from the definition of the out-of-plane wavenumber $w = \sqrt{\eps_{\text{d}/\text{m}} \, \nu_0^2 - u^2}$ and the fact that the squared in-plane wavenumbers $u^2$ exceed $\eps_{\text{d}/\text{m}} \, \nu_0^2$ in each respective medium, with $\nu_0 = E/\left( \hbar \, c \right)$ being the vacuum wavenumber.
Even though the case of infinite media remains a merely theoretical example, it is useful to get a qualitative insight of how the dispersion relation of the SPP at an interface between an opaque metal and a dielectric material (oSPP) can be modified.
In Fig.~\ref{fig_oSPP_disp_shift} the dispersion relation according to Eq.~(\ref{eq_disp_oSPP}) for a high refractive index dielectric medium (average refractive index $n_\text{high} = 1.77$) and a low refractive index medium (average refractive index $n_\text{low}=1.52$) is shown (dashed line).
Hereby, the dielectric function of the metal is modeled by a free electron gas model dielectric function $\eps_\text{m} = 1 - E_\text{p}^2/E^2$.
The plasma energy is fitted to $E_\text{p} = 12.26 \, \text{eV}$ representing aluminum for the relevant range of photon energies.
The reduced refractive index leads to a shift $\Delta u = u_\text{low} - u_\text{high}$ of the oSPP dispersion relation towards smaller in-plane wavenumbers $u$ for all relevant photon energies $E$.
For a photon energy of $ E = 2.6 \,  \text{eV}$, representing the upper bound of interest, the shift is quantified to $\Delta u = - 3.59 \, \mu \text{m}^{-1}$, whereas for the lower energy of $ E = 1.9 \,  \text{eV}$ a smaller shift of $\Delta u = - 2.28 \, \mu \text{m}^{-1}$ is achieved.
From the in-plane wavenumbers, the decay lengths\cite{Maier,DavisII} $\hat z = \left| 1/w_\text{diel} \right|$ within the dielectric can be calculated.
It is found to be in the range of the extension of the optical microcavity.\cite{Maier}
To quantify the shift of the oSPP mode, along with the predictions from the simplified example, Fig.~\ref{fig_oSPP_disp_shift} visualizes the calculated resonance positions of the oSPPs of device A and B.
These involve all the interactions of the complete OLED optical microcavity, calculated from the transfer-matrix.\cite{DavisII,2005_Revelli_wg_analysis_OLED_modes_and_gratings,1992_Smith_Determination_mode_planar_WG_Four_complex_sheets}
Although the positions and absolute values of the shift of the resonance originating from the oSPP, $\Delta u_{2.6\,\text{eV}} = -2.42 \, \mu\text{m}^{-1}$ and $\Delta u_{1.9\,\text{eV}} = - 1.02 \, \mu\text{m}^{-1}$ differ from the predictions derived from Eq.~(\ref{eq_disp_oSPP}), the general trend for the classical case holds, and the oSPP resonance is shifted towards smaller in-plane wavenumbers as the refractive index of the adjacent dielectric is decreased.
The change in the absolute positions of the oSPP and in the slopes of the oSPP dispersions calculated from the transfer-matrix approach compared to the classical model of Eq.~(\ref{eq_disp_oSPP}) arises due to the use of the dielectric functions of all media involved, rather than average refractive indices of the dielectrics.

\subsection{Implications of the oSPP shift on the power dissipation of OLEDs}
The electromagnetic radiation of OLEDs is commonly modeled by the resonant coupling of electric dipole sources to the radiation within the optical microcavity.\cite{Barnes,1998_Benisty_source_term_dipole_emission_in_planar_structures,Webster,FurnoII}
Following this approach, the density of dissipated power by the emitter $K$ for a given photon energy $E$ and in-plane wavenumber $u$ can be calculated.\cite{Webster,FurnoII}
The outcoupled part of the dissipated power
\begin{equation}
F_\text{out} = 2/\nu_\text{active}^2 \int\limits_0^{u_\text{TIR}} \text{d}u \, u \, K_\text{out}(u) \, ,
\end{equation}
 is limited by the in-plane wavenumber $u_\text{TIR} = \sin \left( \theta_\text{TIR} \right) \, \nu_\text{active}$.
This in-plane wavenumber corresponds to the emission angle $\theta_\text{TIR} = \text{arcsin}\left( n_\text{air} / n_\text{active} \right)$ representing trapped light within the optical microcavity due to TIR with respect to the outcoupling medium, i.e. air.
Here $\nu_\text{active} = \nu_0 \, n_\text{active}$ describes the wavenumber within the active medium of the OLED with refractive index $n_\text{active}$.
The outcoupling efficiency $\out{}$ of an optical microcavity is given by the outcoupled power $F_\text{out}$ divided by the total dissipated power
\begin{equation}
F = 2/\nu_\text{active}^2 \int\limits_0^{\infty} \text{d}u \, u \, K(u) \, .
\end{equation}
Thus, the total power contains (i) the outcoupled radiation, i.e. $u < u_\text{TIR}$, (ii) resonance contributions from modes which are propagating in the organic layers, but which are trapped in the device due to TIR, i.e. $u_\text{TIR} < u < \nu_\text{active}$, and (iii) shares of the resonances from radiation represented by plane waves with imaginary out-of-plane wavenumbers $w_\text{active}$, i.e. $u > \nu_\text{active}$ such as SPPs.
Keeping in mind that the dissipated power is calculated from the resonant coupling of the electromagnetic field to the sources, it becomes clear that for evanescent radiation the power dissipation asymptotically vanishes for $u \rightarrow \infty$, since in this limit the electromagnetic field at the emitter position decreases exponentially, and the resonant coupling of the sources to the field decreases likewise.
Furthermore, enhanced dissipation of power occurs into resonances originating from WG and SPP modes, as for these modes resonances of the transmission exist.

By shifting the oSPP towards smaller in-plane wavenumbers, the coupling of the transition dipoles to evanescent modes with $u > u_\text{oSPP}$ is reduced as the asymptotic behavior for $u \gg \nu_0 \, n_\text{active}$  is reached earlier compared to a geometry with an oSPP resonance of a high average refractive index.
\begin{figure*}
\centering
\begin{minipage}{\textwidth}
\includegraphics[width=\textwidth]{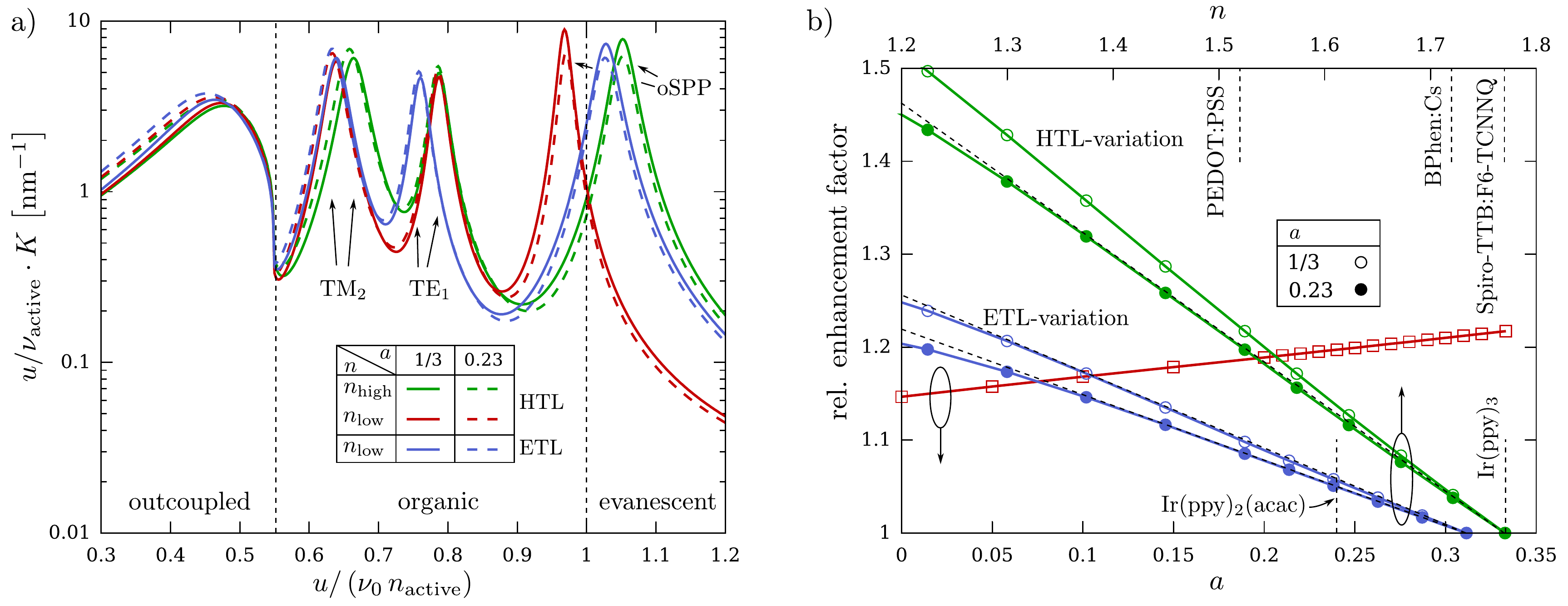}
\end{minipage}
\caption{a) Power dissipation density $u/\nu_\text{active} \cdot F$ of the optimized device A and C comprising the high refractive index HTL (green lines), and B and D using the low refractive index (red lines) HTL material at 510\,nm wavelength.
Simulations are shown for the isotropic emitter ($a=0.3\overline{3}$, solid lines, devices A and B), and anisotropic dipole distributions ($a=0.23$, dashed lines, devices C and D).
A shift of the contribution to the power dissipation spectrum originating from the oSPP causing an increase in outcoupling efficiency by a factor of $\sim 1.2$ for the low average refractive index $n_\text{low}$.
Additionally, a variation of the ETL (blue) is shown where the oSPP shift is significantly smaller compared to the HTL variation.
b) Calculated enhancement factor of the optimized outcoupling efficiencies $\out{}$ with regards to the anisotropy factor $a$ of the emitter (red squares).
The enhancement is obtained for optimized geometries using $n_\text{high}$ and $n_\text{low}$ for the HTL.
Furthermore, the enhancement as a function of the average refractive index $n$ of the HTL (green) or ETL (blue) compared to $n_\text{high}$ is shown for the anisotropy coefficients used in the experiments.
The enhancement factors show only a weak dependence on the anisotropy factor, but they depend heavily on the refractive index difference between $n_\text{high}$ and $n$.
}\label{fig_power_diss}
\end{figure*}
This effect is shown in Fig.~\ref{fig_power_diss}\,a, where the power dissipation spectra of devices A to D are calculated for the refractive indices $n_\text{high}$ (green) and $n_\text{low}$ (red) of the HTL materials, respectively.
The power dissipation is shown for the photon energy  corresponding to the most prominent wavelength $\lambda = 510\,\text{nm}$ of the green emitter materials.
Two anisotropy coefficients\cite{Schmidt} are investigated, with $a=0.3\overline{3}$ (solid lines) corresponding to isotropically distributed transition dipole moments, i.e. Ir(ppy)$_3$, and $a=0.23$ (dashed lines) corresponding to Ir(ppy)$_2$(acac).\cite{2012_Liehm_orientation_APL_Ir3_Ir2_green}
Figure~\ref{fig_power_diss}\,a displays three strong resonances, of which two are located in the organic region for all anisotropy coefficients, representing resonances of WG modes.
The resonance located at $u/\nu_\text{active} = 0.65$ is the excitation of the p-polarized (TM) second order resonance, and the peak at $u/\nu_\text{active} = 0.79$ corresponds to the s-polarized (TE) first order WG excitation.
The third resonance represents the oSPP excitation that is shifted to a value of $u/\nu_\text{active} = 0.96$.
In the region corresponding to the outcoupled radiation, the power dissipation spectrum is defined by shallow resonances which will form the TM$_3$ and TE$_2$ within the organic WG region at higher energies.

Due to the shift of the oSPP the power dissipation into the evanescent region for the lower refractive index HTL is reduced.
Therefore, also the total dissipated power $F$ is strongly reduced by 16.8\,\% for the isotropic emitter and by 15.7\,\% for the anisotropic emitter reducing the refractive index of the HTL.
Additionally, for this microcavity the outcoupled power $F_\text{out}$ is slightly increased due to the enhanced emission into resonant modes of the microcavity, as can be seen from Fig.~\ref{fig_power_diss}\,a.
For $a=0.3\overline{3}$ the relative increase of the outcoupled power is quantified to be 5.0\,\% and for $a=0.23$ by 2.9\,\%.
Hence, in total the outcoupling efficiency is increased by 22\,\% for $a=0.3\overline{3}$, resp. 18\,\% for $a=0.23$.
Thus, it follows that the majority of the increase is achieved by the displacement of the oSPP.

Furthermore, Fig.~\ref{fig_power_diss}\,a shows the optimized power dissipation of a comparable refractive index variation of the ETL of devices A and C (blue).
Due to the extent of the oSPP electromagnetic field over the complete optical microcavity, the reduction of the refractive index of the ETL also leads to a shift of the oSPP dispersion relation.
But compared to the HTL variation, in this case the shift is much weaker due to the increased distance between the low refractive index medium and the surface of the opaque metal bottom reflector.
In contrast, the intensity and position of the higher order WG modes (TE$_1$, TM$_2$, and the radiative TM and TE modes) is modified.
This arises due to the fact, that the electromagnetic field of these modes is much more localized around the emission layer and ETL.
Thus, the reduced refractive index of the ETL has a larger influence on the WG resonance positions and intensities than for a modified HTL.
In summary, the outcoupling efficiency is enhanced by about 9.8\,\% for isotropic dipole distributions for a refractive index change of about 12\,\% of the ETL, and by 8.5\,\% for oriented dipole distributions with an anisotropy factor $a=0.23$.
Thus, the usage of the low refractive index HTL material close to the opaque back reflector can influence the oSPP two times more efficiently compared to a low refractive index ETL. 

To substantiate the potential of the proposed method Fig.~\ref{fig_power_diss}\,b shows the calculated enhancement factors for $\out{}$ in dependence of the anisotropy factor (red) and the refractive index of the HTL (green) and ETL (blue).
For the variation of the anisotropy coefficient the refractive index of the HTL was modified from $n_\text{high}$ to $n_\text{low}$.
The outcoupling efficiencies were obtained from the simulation of the optimized device geometry.
For the entire range of investigated anisotropy coefficients and enhancement of the optimized outcoupling efficiency is observed.
Even for completely in-plane oriented dipole sources the enhancements would be about around 15\,\%, for a refractive index $n_\text{low}$ reduced by about 14\,\% to the reference value $n_\text{high}$.
This is caused by the fact that the emission into SPPs for emitter materials aligned in-plane is reduced compared to isotropic emitters.
However, the oSPP contribution does not vanish due to the interaction of p-polarized radiation with the horizontally aligned dipoles.
Hence, the total dissipated power is still reduced for the complete range of anisotropy factors.
Regardless of the alignment of the dipole distribution the increased intensities of the higher order WG modes holds, causing an increased outcoupled power, resulting in an overall enhancement.
Moreover, for a range of anisotropy coefficients covered by existing phosphorescent emitter materials, i.e. $a \in \left[ 0.23,\,0.3\overline{3} \right]$, the enhancement remains around 20\,\% and arises mainly from the shift of the oSPP mode.
But also for $a=0.08$\cite{2014_Mayr_Emitter_0.08_orientation_TADF}, recently found in a thermally activated fluorescent (TADF) emitter material, the enhancement would still be about 17\,\%.

Furthermore, Fig.~\ref{fig_power_diss}\,b reports the enhancement factors of the optimized outcoupling efficiencies as a function of the average refractive index of the HTL (green) and ETL (blue) of optimized first order OLED microcavities.
The refractive indices are varied for device geometries incorporating transition dipole orientations currently realized by phosphorescent emitters.
For average refractive indices smaller than the reference value of $n_\text{high}$, the enhancement increases approximately linearly decreasing the average refractive index for both the HTL and ETL materials.
The enhancement factors for average refractive indices below 1.3 are considered to be hypothetical, as materials with such low refractive indices and reasonable conductivities are hard to obtain.
Nevertheless, this figure indicates that at very low refractive indices the linearity is dropped and the enhancement factor will saturate.
This is caused by the shift of the oSPP, which also saturates towards average refractive indices $n = n_\text{air}$.
Likewise, it can be seen that the oSPP shift is the predominant cause for the enhanced outcoupling efficiency comparing the enhancements of the HTL variations to the ETL alteration.
Hence, it is more desirable to modify the refractive index of the HTL than the index of the ETL, albeit a combination of both would lead close to the high predicted efficiencies by Smith and Barnes.\cite{Smith}

\subsection{Optimizing thin film SPPs and power dissipation of implemented devices}
The full benefit of our optimization strategy arises (for the top-emitting geometries) if the excitation of resonant modes with $u > u_\text{oSPP}$ can be avoided entirely.
For top-emitting OLED optical microcavities, these modes consist of SPPs excited at the thin semi-transparent top-contact as long as this contact contains any metal.
The SPPs of thin metal films are the coupled states of the SPPs excited at each metal/dielectric interface, where they are able to form a symmetric (sSPP) and an anti-symmetric (aSPP) state, regarding the z-component of the electrical field for a symmetric $\eps_1/\eps_\text{met}/\eps_1$ geometry.
For a non-symmetric geometry $\eps_1/\eps_\text{met}/\eps_3$ the sSPP describes a mode with a minimum of the z-component of the electrical field within the metal, and the aSPP a mode with a field envelope changing sign inside the thin metal layer.
In this simple case the dispersion relation can be calculated numerically from\cite{Raether,Maier}
\begin{equation}\label{eq_thin_SPP}
\mate^{2 \, \imath \, \delta \, w_\text{met}} = \frac{\left( w_\text{met} \, \eps_1 + w_1 \, \eps_\text{met} \right) \, \left( w_3 \, \eps_\text{met} + w_\text{met} \, \eps_3 \right)}{\left( w_\text{met} \, \eps_1 - w_1 \, \eps_\text{met} \right) \, \left( - w_3 \, \eps_\text{met} + w_\text{met} \, \eps_3 \right)} \, ,
\end{equation}
where $\delta$ denotes the thickness of the metal layer.
Calculating the in-plane wavenumbers for the sSPP and the aSPP from Eq.~(\ref{eq_thin_SPP}) or the transfer-matrix of a complete OLED geometry, for common metals one obtains, $\Re\left(u_\text{sSPP}\right) < \Re\left(u_\text{oSPP}\right) < \Re\left(u_\text{aSPP}\right)$ for the real parts of the in-plane wavenumbers within the visible spectral range.
However, due to the strong in-plane localization of the sSPP, this mode is irrelevant for the power dissipation in OLED geometries.\cite{DavisII,2012_Furno_SPP_in_org_microcav}
In contrast, the aSPP can be excited from the emissive dipoles, similar to the oSPP, although the excitation is weaker due to the larger in-plane wavenumber $u_\text{aSPP}$.

\begin{figure*}
\centering
\begin{minipage}{\textwidth}
\includegraphics[width=\textwidth]{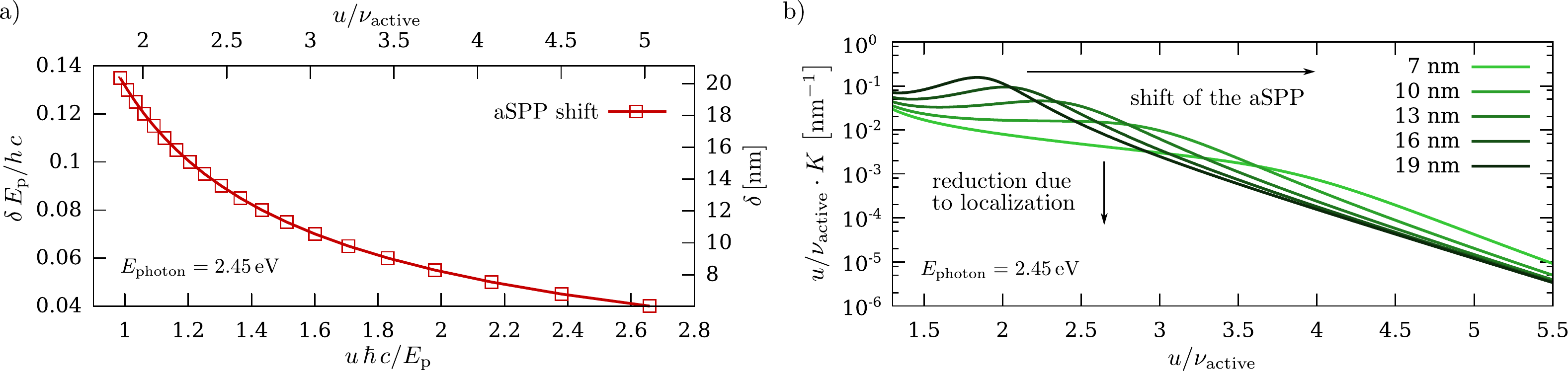}
\end{minipage}
\caption{a) Calculated shift of the aSPP mode, from Eq.~(\ref{eq_thin_SPP}), for various thicknesses of the metal layer. The dielectric functions $\eps_1 = 2.96$ and $\eps_3 = 3.53$ correspond to BPhen:Cs and $\alpha$-NPD respectively. The dielectric function of the metal was parametrized as a Drude model with $E_\text{p} = 8.22$\,eV, corresponding to silver.
A non-linear increasing shift of the aSPP position decreasing the metal thickness $\delta$ is observed.
b) Simulated power dissipation spectra for a photon energy of 2.45\,eV taking into account the complete OLED stack (device A) and varying the silver top contact thicknesses $\delta$.
Due to the aSPP shift the dissipated power into this mode is strongly reduced.
}\label{fig_thin_film_SPP}
\end{figure*}

In Fig.~\ref{fig_thin_film_SPP}\,a the in-plane wavenumbers for the aSPP mode at a fixed photon energy of $E_\text{photon} = 2.45\,\text{eV}$ is calculated for for decreasing metal thicknesses $\delta$.
The dielectric functions $\eps_1 = 2.96$ and $\eps_3 = 3.53$ correspond to the ETL material BPhen:Cs and the CL material $\alpha$-NPD at this energy.
The plasma energy $E_\text{p}$ in the Drude model dielectric function of the metal was fitted to $E_\text{p} = 8.22\,\text{eV}$, representing the silver forming the thin film electrode.
A nonlinear shift of the aSPP position as a function of the metal layer thickness $\delta$ is observed, with increasing shifts for decreasing thicknesses $\delta$.

For decreased metal film thicknesses $\delta$, the nonlinear shift of the aSPPs in-plane wavenumber towards higher values causes a pronounced localization of the aSPPs electromagnetic field in and around the metal film.
Thus, the interaction with the emitter sources decreases.
Hence, the power dissipation into the aSPP is reduced for decreased metal film thicknesses.
This shift of the aSPP and the reduced power dissipation into the resonance is shown in Fig.~\ref{fig_thin_film_SPP}\,b.
Here the displacement of the aSPP excitation of the complete OLED microcavity is in good agreement with the predictions from Eq.~(\ref{eq_thin_SPP}) (cf. Fig.~\ref{fig_thin_film_SPP}\,a) taking into account the dielectric function of the top electrode silver and the additional thin gold film.
These simulations included all layers of the optical microcavity of device A with fixed thicknesses, while the thickness of the semi-transparent silver layer was varied.
Due to the decreased contribution of the aSPP and the broadening of the spectral range of high transmission, the outcoupling efficiency of the OLED optical microcavity is enhanced.
This holds for the high refractive index HTL material, $\out{,\,\text{high},\,7\,\text{nm}}/\out{,\,\text{high},\,20\,\text{nm}} = 1.11 $, as well as for the low refractive index HTL, $\out{,\,\text{low},\,7\,\text{nm}}/\out{,\,\text{low},\,20\,\text{nm}} = 1.12 $ using isotropic dipole distributions.
But also for the anisotropic distribution $a=0.23$, $\out{,\,\text{high},\,7\,\text{nm}}/\out{,\,\text{high},\,20\,\text{nm}} = 1.10 $ and $\out{,\,\text{low},\,7\,\text{nm}}/\out{,\,\text{low},\,20\,\text{nm}} = 1.11 $ are obtained.

Within the experiment, such ultra-thin metal layers are realized by using a recently adapted wetting layer electrode approach\cite{2013_Schubert_ultra_thin_metal_seed,2013_Schwab_OLED_wetting_layer_electrode}, where a seed metal is used to form a closed conductive and highly transparent metal electrode.
With this approach, the combined thickness of the smooth metal electrode does not exceed 10\,nm, enabling very high outcoupling efficiencies.
\begin{figure*}
\centering
\begin{minipage}{\textwidth}
\centering
\includegraphics[width=\textwidth]{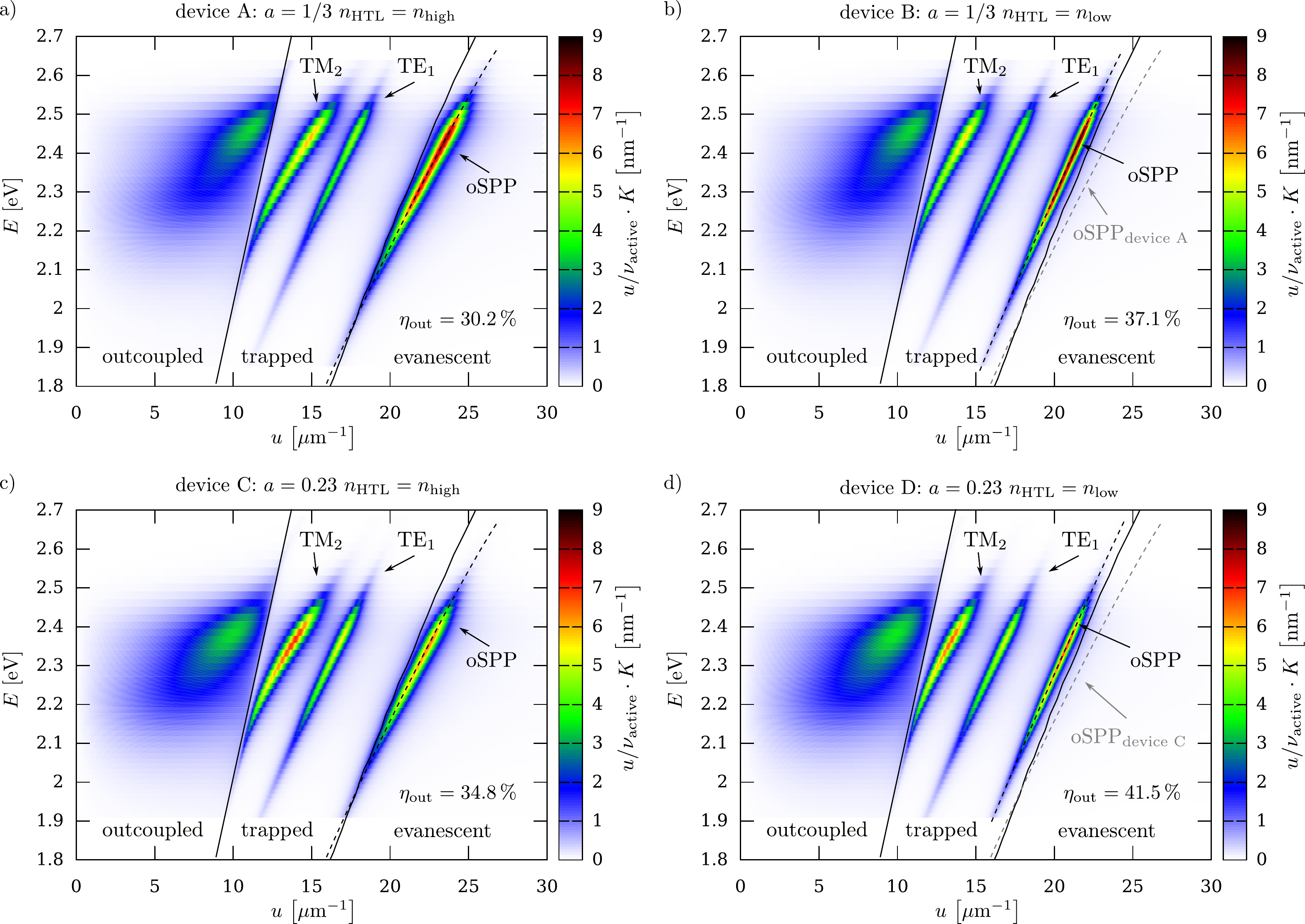}
\end{minipage}
\caption{a-d) Calculated power dissipation spectra for the relevant spectral range for devices A-D.
The air light line and the organic light line separate the outcoupled from the trapped and the trapped radiation from the evanescent excitations, respectively (black solid lines).
The contributions to the power dissipation originating from the oSPP resonances are highlighted as dashed lines.
Comparing devices A and B to C and D the optical thickness of the microcavities is maintained while reducing the refractive index of the HTL
Thus, the contributions from WG resonances are only slightly modified.
The oSPP contribution is reduced for devices C and D due to the anisotropic emitter orientation.
Hence, the outcoupling efficiencies of C and D exceed those of devices A and B.
However, the enhancements due to the oSPP shift is comparable with $\out{,\,\text{B}}/\out{,\,\text{A}}= 1.23$ and $\out{,\,\text{D}}/\out{,\,\text{C}}= 1.19$.}\label{fig_power_diss_all}
\end{figure*}

The modification of the power dissipation spectrum and the improved outcoupling efficiency are shown in Fig.~\ref{fig_power_diss_all}.
It reports the power dissipation spectra over the complete spectral range for Ir(ppy)$_3$ (devices A and B) or Ir(ppy)$_2$(acac) emitters (devices C and D).
The air light lines and the organic light lines, defined from the refractive index of the active emitter medium, are superimposed to the contour plots.
These lines distinguish between the power outcoupled into air, power dissipated into trapped light propagating as WG modes within the organic, and the power dissipated into evanescent excitations lost as heat.
Decreasing the refractive index of the HTL material (Fig.~\ref{fig_power_diss_all}\,b and d) shifts the oSPP resonance to in-plane wavenumbers below the light line of the organic medium.
For devices A and B, the displacements correspond to the ones given in the discussion of Fig.~\ref{fig_oSPP_disp_shift}.
The shifts for the devices C and D are given by $\Delta u_{2.6\,\text{eV}} = -2.19\,\mu\text{m}^{-1}$ and $\Delta u_{1.9\,\text{eV}} = -0.93\,\mu\text{m}^{-1}$.
Compared to A and B, the smaller shift of the oSPP for devices C and D is attributed to the reduced HTL thickness caused by the anisotropy allowing to reduce the optimized distance between the emitter and the opaque metal electrode.
This is also reflected by the calculated efficiencies which are $\out{,\,\text{A}} = 30.2\,\%$, $\out{,\,\text{B}} = 37.1\,\%$, $\out{,\,\text{C}} = 34.8\,\%$, and $\out{,\,\text{D}} = 41.5\,\%$.
The relative enhancement of the outcoupling efficiency of device B compared to A is about 1.23, whereas the relative enhancement for device D compared to C is 1.19.
The reduction is attributed to the smaller shift of the oSPP mode resonance together with an decreased impact of the oSPP on the total dissipated power for anisotropic emitter materials, cf. Fig.~\ref{fig_power_diss}\,b.
However, for both emitter materials, the oSPP resonance shifts to in-plane wavenumbers $u_\text{oSPP} < \nu_\text{active}$ over the complete spectral range, where the in-plane wavenumbers of the resonances correspond to internal emission angles of about $\theta \approx 75^\circ$.
Hence, this resonance can be addressed by macroscopic extraction structures\cite{2007_Yang_top_OLEDs_w_structure_4_outcoupling,Thomschke2012}, as within the organic semiconductor materials the oSPP is represented by a non-evanescent propagating plane wave.
Furthermore it is worth to note that due to the oSPP shift, the excited modes are compressed into a smaller range of reciprocal space.
This implies higher enhancement of such top-emitting OLEDs if Bragg scattering structures\cite{Matterson,2004_Smith_top_emitting_OLED_simulation_PD,Bi,2014_Schwab_Fuchs_Opt_Exp} are applied, because for one reciprocal lattice constant more modes can be scattered into the outcoupling cone.
Moreover, as the in-plane wavenumber of the oSPP is reduced, the lattice constants for which the plasmon contribution is addressed can be increased compared to standard HTL materials.
Having outlined the theoretical aspects of the optimization concept, the experimental realization is discussed in the next section.

\section{Experimental realization and results}
Devices A to D were fabricated from the materials given in Sec.~\ref{sec_materials}.
All materials, except the PEDOT:PSS were deposited by thermal evaporation in an UHV tool (Kurt J. Lesker Co.) at a base pressure of 10$^{-7}$--10$^{-8}$\,mbar.
The PEDOT:PSS HTL was spin coated onto the opaque bottom electrode adjusting the speed of rotation to obtain the desired thicknesses followed by a 15\,min baking step at 110\,$^\circ$C under ambient condition.
To enhance the wetting behavior of the PEDOT:PSS solution onto the aluminum electrode, and thus to achieve homogeneous transparent films, the substrates containing the electrodes were exposed to an argon plasma at a base pressure of 10$^{-2}$\,mbar for ten minutes.
In order to obtain comparable electrodes for all devices, the plasma treatment was applied to all electrodes, including the subsequent exposure to ambient conditions.
Prior to the EBL deposition under UHV conditions, all substrates were transferred to vacuum and again heated out for 1\,h at 110$^\circ$\,C.
Directly after the last deposition step, the devices were encapsulated with a glass lid under nitrogen atmosphere.
A source measure unit (Keithley 2400) and a calibrated spectrometer (Instrument Systems CAS140) providing the spectral radiant intensity were used to obtain the current-voltage-luminance ($I$-$V$-$L$) characteristics of all devices.
The EQEs are calculated from the $I$-$V$-$L$ and the angle-dependent emission spectra of the OLEDs, which were obtained for emission angles $\theta$ ($0^\circ \le \theta \le 90^\circ$) with a spectro-goniometer using a calibrated miniature spectrometer (USB 4000, Ocean Optics Co.).
\begin{figure*}
\centering
\begin{minipage}{\textwidth}
\centering
\includegraphics[width=0.7\textwidth]{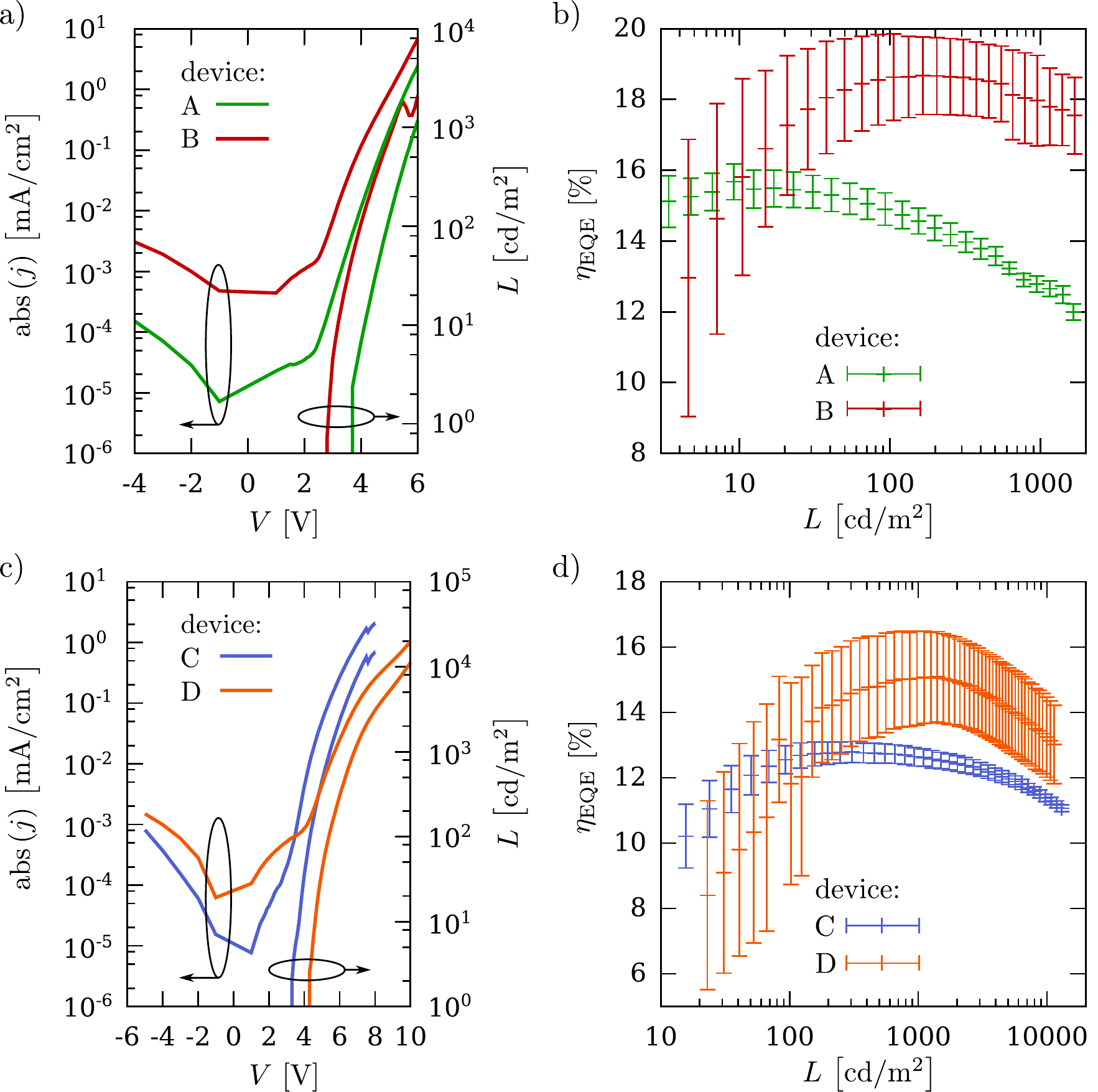}
\end{minipage}
\caption{a) Measured $j$-$V$-$L$ characteristics for devices A and B.
Due to the use of spin-coated PEDOT:PSS the leakage currents are increased, and the slope of the $j$-$V$ curve is decreased beyond the threshold voltage.
b) External quantum efficiencies $\EQE{}$ obtained from the measured angle dependent emission spectra of devices A and B.
The $\EQE{}$ of the device B is increased by a factor of $1.19$ compared to device A, in agreement with the value of 1.23 predicted by the optical simulations.
c) Measured $j$-$V$-$L$ characteristics for devices C and D.
Similar effects due to the PEDOT:PSS occur for device D as for device B.
d) EQEs derived from the measurements on devices C and D.
The maximum EQE is enhanced by about 1.18, again in good agreement with the optical enhancement value of 1.19.
The errors of the EQEs for devices B and D are increased due to the increased sample-to-sample variations caused by spin coating PEDOT:PSS.}\label{fig_IVL_EQE}
\end{figure*}
The $j$-$V$-$L$ characteristics for devices A and B are given in Fig.~\ref{fig_IVL_EQE}\,a and for C and D in Fig.~\ref{fig_IVL_EQE}\,c, where $j$ denotes the current density from which the current can be calculated implying an active area of the OLED of 6.76\,mm$^2$.
Both OLEDs using the PEDOT:PSS HTL exhibit an increase in the leakage current by about one order of magnitude compared to their reference devices.
We attribute this effect to the \textit{ex situ} deposition of PEDOT:PSS, as this process is likely to introduce impurities into the sensitive organic layers.
Furthermore, we observe a decreased slope of the $j$-$V$ curve beyond the threshold voltage for the low refractive index HTL devices.
By changing the HTL material from Spiro-TTB:F6-TCNNQ to PEDOT:PSS, the charge injection barriers from the anode to the HTL \cite{2003_Joensson_PEDOT_AL_Interaction_deep_depletion} and from the HTL to the EBL\cite{2010_Yamanari_PEDOT_acid_interface_mess} are modified.
Especially due to the corrosivity of the aqueous PEDOT:PSS dispersion\cite{2010_Yamanari_PEDOT_acid_interface_mess} the formation of capacitive interlayers on the anode surface is enhanced\cite{2003_Joensson_PEDOT_AL_Interaction_deep_depletion}, introducing additional barriers for electrical transport, and thus changing the $j$-$V$ characteristics.

Figure~\ref{fig_IVL_EQE}\,b and d show the EQEs measured for all four devices.
The errors, derived from the sample-to-sample variation from eight measured OLEDs for each device, of the EQEs for devices B and D are increased compared to the reference devices A and B.
This is explained by the spin coating process for the PEDOT:PSS HTL, because this process induces sample to sample variations of the film thickness and introduces more defects on the OLEDs, as discussed earlier.
For the maxima of the $\EQE{}$ the charge carrier balance $\gamma$ must also be maximized, if the maximum occurs at luminance where non-radiative roll-off processes can be neglected.
Since PEDOT:PSS is also a doped hole transport material with comparable conductivity to Spiro-TTB, it is assumed that at the maximum $\gamma$ is comparable for both HTLs.
However, due to this change in the electrical characteristics the maximum is likely to occur at different current densities, i.e. luminance, using the PEDOT:PSS HTL for devices B and D.
Hence, instead of comparing EQEs at fixed luminance, the maxima of the $\EQE{}$ will be compared for the corresponding pairs of devices.
Thus, the increase in the maximum of the EQE is attributed to the enhancement originating from the outcoupling efficiency.
We find maximized EQEs of $\EQE{,\,\text{A}} = 15.7 \, \%$ and $\EQE{,\,\text{B}} = 18.7 \, \%$ for devices A and B, and $\EQE{,\,\text{C}} = 12.8 \, \%$ and $\EQE{,\,\text{D}} = 15.2 \, \%$ for devices C and D.
This corresponds to enhancements of the maximum EQEs by factors of $\EQE{,\,\text{B}}/\EQE{,\,\text{A}} = 1.19 $ and $\EQE{,\,\text{D}}/\EQE{,\,\text{C}} = 1.18 $, in good agreement with the predicted enhancements of the outcoupling efficiencies $\out{,\,\text{B}}/\out{,\,\text{A}} = 1.23 $ and $\out{,\,\text{D}}/\out{,\,\text{C}} = 1.19 $.
The increased EQEs for devices A and C for small luminance is thus attributed to the increased leakage currents for the PEDOT:PSS devices and the decreased charge carrier balance $\gamma$.
Furthermore, we attribute the decreased EQEs for devices C and D, compared to the devices A and B, to an increased sensitivity of the Ir(ppy)$_2$(acac) emitter molecule to the deposition conditions of the anode in combination with the PEDOT:PSS.

\section{Conclusions}
In this paper, we have proposed a method to enhance the outcoupling efficiency, and thus the EQE, of monochrome green top-emitting OLEDs by modifying the oSPP resonance of the optical microcavity.
The modification of the oSPP was achieved by using an HTL on top of the opaque electrode of the OLED, with a reduced average refractive index compared to a state-of.the-art HTL.
The simulations have predicted a linear increase of the relative outcoupling efficiency enhancement proportional to the relative change of the average refractive index.
According to these simulations, the outcoupling efficiency increases by about 20\,\% when using a reduction of the average refractive index by 14\,\%.
This enhancement holds for a wide range of anisotropy parameters of organic emitter molecules.
Even for completely horizontally aligned transition dipole moments, the enhancement would still be about 15\,\% if ultra thin top contacts are used.

We have verified our results from optical simulations by fabricating top-emitting green phosphorescent OLEDs, using the isotropic emitter Ir(ppy)$_3$ and the anisotropic emitter Ir(ppy)$_2$(acac).
In both cases we find relative enhancement factors for the maximum of the EQEs: 1.19 for Ir(ppy)$_3$ and 1.18 for Ir(ppy)$_2$(acac), in good agreement with the predictions based on the optical simulations.

Furthermore, we have found that the combination of the opaque aluminum anode and the proposed low refractive index HTL produces an oSPP resonance propagating as a non-evanescent plane wave within the layers of the optical microcavity, excluding the HTL itself and the metal layers.
Thus, this resonance becomes suitable for further macroscopic outcoupling techniques applicable to top-emitting OLEDs.\cite{2007_Yang_top_OLEDs_w_structure_4_outcoupling,Thomschke2012}
Besides, coherent outcoupling techniques, such as Bragg scattering\cite{Bi,2014_Schwab_Fuchs_Opt_Exp}, should be more efficient for these types of optical microcavity, because the phase space into which modes are excited is compressed to a more narrow range of in-plane wavenumbers.
Furthermore, the modification of the oSPP can be obtained for all photons energies within the visible spectrum.
Hence, using low refractive index HTL materials may be suitable to enhance the performance of white light top-emitting devices as well.

\acknowledgments 
This work was funded by the European Social Fund and the free state of Saxony through the OrganoMechanics project.
Support from the excellence cluster cfaed is gratefully acknowledged.


\bibliography{new_lit}   

\end{document}